# Where to Refuel: Modeling On-the-way Choice of Convenience Outlet

**Ari Pramono**

Department of Marketing

MONASH BUSINESS SCHOOL

**Harmen Oppewal***

Department of Marketing

MONASH BUSINESS SCHOOL

Ari Pramono, PhD, is a Research Fellow in the Department of Marketing of the Monash Business School, Monash University, Caulfield East, VIC 3145, Australia, email: Ari.Pramono@monash.edu;

Harmen Oppewal, PhD, is Professor of Marketing, Department of Marketing, Monash Business School, Monash University, 26 Sir John Monash Drive, Caulfield East, VIC 3145, Australia, email: Harmen.Oppewal@monash.edu, tel. +61 3 9903 2360.

*) Corresponding author

**WHERE TO REFUEL:**

**MODELING ON-THE-WAY CHOICE OF CONVENIENCE OUTLET**

ABSTRACT

For many goods consumers do not make a special trip to a store. Especially for a convenience good such as fuel they will buy the product while on-the-way to some final destination. This paper introduces on-the-way choice of retail outlet as a form of convenience shopping. It presents a model of on-the-way choice of retail outlet and applies the model in the context of fuel retailing to explore its implications for segmentation and spatial competition. The model allows analyzing how choice of retail outlet varies not only with spatio-temporal variables (distance, detour, local competition and agglomeration) but also with trip-related characteristics such as time of day and prior awareness of one's purchase need. The model is a latent class random utility choice model. An application to gas station choices observed in a medium-sized Asian city show the model to fit substantially better than existing models. The empirical results indicate consumers may adopt one of two decision strategies. When adopting an immediacy-oriented strategy they behave in accordance with the traditional gravity-based retail models and tend to choose the most spatially convenient outlet. When following a destination-oriented strategy they focus more on maintaining their overall trip efficiency and so will tend to visit outlets located closer to their main destination and are more susceptible to retail agglomeration effects. The paper demonstrates how the model can be used to inform segmentation and local competition analyses that account for variations in these strategies as well as variations in consumer type, origin and time of travel. Simulations of a duopoly setting further demonstrate the implications.

Keywords: Spatial choice modeling; Convenience retailing; Retail location; Mobile consumption; Fuel markets

Changes in consumer lifestyles, increased mobility and the rise of the 24/7 economy have made consumers increasingly time poor and reliant on convenience shopping to make many of their purchases (Mintel 2018). A particular type of convenience shopping exists when consumers need to visit a retail outlet while they are on their way to a main destination. Products for which this applies include on-the-go purchases of food and drinks, purchases of magazines or batteries, and cash withdrawals. Another significant example is when motorists need to refuel their car. In all these cases, the time and location where the purchase need arises, the directionality of the travel to the main destination and the type of trip can be expected to influence the attractiveness, and therefore the competitiveness, of a retail outlet for purchasing the item. In particular, as this paper will argue, the attractiveness of an on-the-way retail outlet not only depends on its direct distance to the consumer but also on its location relative to the consumer's final destination and the time and purpose of traveling.

Despite its relevance, the literature provides only limited insight into on-the-way store choice. Convenience relates to reductions in consumers' effort and time spend (Berry, Seiders and Grewal 2002). The retail literature has studied consumer store choice in various convenience-related contexts, including fill-in trips (Kahn and Schmittlein 1989; Reutterer and Teller 2009) and multipurpose shopping studies (Arentze, Oppewal and Timmermans 2005; Dellaert et al. 1998; O'Kelly 1981; Popkowski-Leszczyc, Sinha and Sahgal 2004; Thill 1992) but none of these take on-the-way contexts into account. Early literature identifies convenience goods as goods that consumers usually purchase at easily accessible locations (Copeland 1923) and as goods where consumers obtain a lower gain from price and quality comparisons when searching the product (Holton 1958). The literature also identifies that consumers typically do not make a special trip to acquire them (Eaton and Tweedle 2012), although this does not preclude convenience trips from being planned trips. Recently, Blut, Teller and Floh (2018) in a meta-analysis identified ease of access, proximity and distance, which are related to convenience, as key factors of the retail marketing mix. Recent studies also have identified 'on-the-go' consumption, which is the purchase and consumption of food and beverages while the consumer is in transit (Heider and Moeller 2012). Studies of on-the-go consumption have addressed preference heterogeneity and the choice between traditional and on-the-go outlets (Benoit, Evanschitzky and Teller 2019; Benoit, Schaefers and Heider 2016; Sands et al. 2019) but not actual store choice.

Indeed, no prior work on store choice seems to directly account for the time and directionality of on-the-way behavior, or for the situation where the final destination is different from the origin or starting location. This lack of insight not only limits our ability to predict store choices but also limits our understanding of the nature of local competition in this particular context. It is already well understood that spatial substitutability between retail outlets depends on how consumers trade-off spatial and non-spatial characteristics, for example, for different retail formats (Ellickson, Grieco and Khvastunov 2020; Fox, Montgomery and Lodish 2004; Fox and Sethuraman 2010) or between premium/network and low/independent branded gas stations (Eckert and West 2006; Lewis 2008). However, the local variation in spatial substitutability that may arise when consumers are on-the-way is not well understood. The present paper seeks to fill this gap by introducing a model of on-the-way choice of retail outlet. Our aim is twofold: 1) Propose a model that allows to better understand and quantify how various trip-related factors influence on-the-way choice of retail outlet and 2) apply the model to explore its implications for segmentation and spatial competition.

We specifically study where motorists refuel their car. Refueling is a typical on-the-way convenience purchase (Eaton and Tweedle 2012). Motorists will seek to refuel at a location that does not require deviating much from the intended route. This means that stations along the consumer's route are potential substitutes while those nearby but off the main route may not (Houde 2012; Kitamura and Sperling 1987; Rossi and Chintagunta 2016). This substitutability between gas stations will however vary with the heterogeneity in consumers' trip paths (Chandra and Tappata 2011) or demography (Nishida and Remer 2018). Furthermore, as motorists trade off travel time against non-spatial attributes such as price and quality and seek to increase the efficiency of their trip, they may combine their fuel stops with other activities. This means the substitutability will also depend on the extent to which gas stations benefit from being located near other activity locations.

Previous studies on refueling mostly assume that consumers will always choose the nearest alternative (Chan, Padmanabhan and Seetharaman 2007; Eaton and Tweedle 2012; Houde 2012; Iyer and Seetharaman 2008). Houde (2012) relaxes the nearest outlet assumption by allowing consumers to choose gas stations located farther away, as long as they are on the path to the consumer's main destination. This is in accordance with Kelley and Kuby (2013)'s finding that drivers chose ten times more the station that was most convenient on their way than the station closest to their residential location. Houde (2012) however still maintains the assumption that consumers always decide at their home location when and where to refuel. Other studies measure competition between fuel stations from aggregate data such as sales and relate these to observed prices. They find that the competition between gas stations is highly localized (Nishida and Remer 2018) and is driven by variation in non-spatial attractiveness (Lewis 2008). Yet another study identifies the effects of price signs (Rossi and Chintagunta 2016). None of the above studies however involve choice models that allow predicting individual outlet choices and that allow accounting for differences in local competition based on variations in trip characteristics and consumer choice strategies.

Our study observes and analyzes the travel paths of a sample of consumers who were intercepted at their point of purchase at different times during weekdays in a particular urban area. Price and brand were uniformly distributed in our study area; hence, our analysis focuses on the role of spatial variables although it still also includes station quality as a non-spatial attraction variable. We model the outlet choice of the observed consumers by taking into account their specific spatio-temporal settings. As such, our model can be used to predict consumer outlet choices in a variety of on-the-way contexts, thereby providing insight into the factors determining spatio-temporal variation in choice probabilities and, consequently, in the levels of local competition. In order to assess the variations in substitutability between outlets in the area for different times of the day we aggregate model predictions across travel conditions to obtain sample level estimates of substitutability. The model also provides a basis for the identification of new opportunities for segmentation. Potentially relevant market segments can be derived from variations in consumer preferences regarding outlet quality and travel distance but also in terms of the place and time when they can be most effectively reached with market communications, for example promotional offers that aim to increase an outlet's patronage.

In our modelling we in particular allow for two different strategies that consumers can adopt when needing to choose a retail outlet while on-the-way and identify determinants and implications of differences between these strategies. First, consumers can anticipate their need and plan ahead where to service it—we call this a *destination-oriented* strategy. Alternatively, consumers can adopt an *immediacy-based* strategy, which is where they decide only while they are on-the-way where to service their need. This means they will have to rely on whichever provider is locally available and have less opportunity to find the highest quality or lowest price retailer. They will also be less prone to combine their stop with other local services, thus reducing the relevance of agglomeration advantages, which are the benefits a store derives from being located close to a retail outlet from a different category (Arentze et al. 2005; Brown 1989; Teller, Alexander and Floh 2016; Teller and Reutterer 2008).

Our paper contributes to the retail literature in several ways. First, by identifying the on-the-way setting as a specific type of convenience context. As noted, the literature so far has not identified the unique aspects of this context. Second, we provide new insights into how consumer and outlet characteristics influence destination choices, in particular the role of detour, point of need awareness and decision strategy adoption. As such we contribute to the understanding of an increasingly important and complex type of shopping behavior. Third, our analysis reveals effects of the on-the-way shopping context on spatial competition and agglomeration among retail outlets that have remained undiscovered when using other models. Further, our model comprises a spatial choice model that predicts outlet choice in the context of a consumer's wider goal of reaching a main destination. It thus extends work on convenience and multipurpose shopping. Finally, our model can be used to assess and measure spatio-temporal variation in substitutability between outlets. Existing studies in convenience retail have not allowed for this.

Our paper is structured as follows. We first briefly review fuel retailing and spatial choice models and explain why refueling presents the prototypical case of on-the-way choice. We next introduce our model and apply it in a specific urban setting. We next compare the model against benchmark models and traditional retail models. We illustrate how it can assist with segmentation and competitive analysis and demonstrate implications of the two strategies in a series of simulations. We finish with a discussion of the implications of our findings.

*FUEL RETAILING AND SPATIAL MODELS*

The gasoline retail sector represents an important market by itself but also provides a very suitable context for investigating spatial competition (Houde 2012; Paul, Miljkovic and Ipe 2001; Pennerstorfer 2009). It is distinctive in the consumer's mobility while shopping for and consuming the product (Houde 2012) and displays significant variation in pricing (Nishida and Remer 2018). Fuel retailing is also undergoing significant change, with an increasing integration with non-fuel sales and involvement of major supermarkets (Economist 2015, 2019) and the transition to electrical and driverless vehicles (BCG 2019). Modeling fuel outlet choices can help understand how consumers will respond to these changes (Sun, Yamamoto and Morikawa 2016).

We study a fuel market in which many factors are controlled that otherwise limit the analysis of fuel outlet choice. Due to a state monopoly, in our study area there is only a single brand of fuel and all prices are centrally fixed. The stations however vary in their service levels and so, their non-spatial attractiveness still varies. There are no non-fuel items at the gas stations. Our setting thus provides a unique opportunity to study spatial aspects of outlet choice.

Our model accounts for distance deviation (detour), for the location where a purchase need actually arises and for the existence of multiple decision strategies. Current store choice models all consider consumers to be static and located in one place. This includes gravity (Huff 1964; Nakanishi and Cooper 1974; Reilly 1931), spatial interaction (Haynes and Fotheringham 1984; Newing, Clarke and Clarke 2015), competing destination (Fotheringham 1988), and disaggregate spatial choice models (Albuquerque and Bronnenberg 2012; Arentze et al. 2005; Weisbrod, Parcells and Kern 1984). However, as fuel and convenience customers move across the market when consuming the product, their position relative to specific choice alternatives shifts and the convenience of accessing the alternatives changes (Claycombe 1991; Houde 2012).

Although existing methods account for clustering and agglomeration effects (Arentze et al. 2005; Borgers and Timmermans 1987; Fotheringham 1988) and neighborhood adjacency of alternatives (Guo and Bhat 2004; Sener, Pendyala and Bhat 2011), they do not account for the location where the need to refuel arises, nor for the spatial integration of the refueling activity with other activities or for route familiarity. Path-based models (Hui, Bradlow and Fader 2009) account for spatial configuration effects but only in limited ways. While there are similarities with multi-category purchase models (Seetharaman et al. 2005), these neither account for spatiality. Multipurpose shopping trip models (Arentze et al. 2005; Dellaert et al. 1998; Gijsbrechts, Campo and Nisol 2008; Popkowski-Leszczyc et al. 2004) accommodate that a purchase may be made in a farther away location because it can be combined with another purchase at that same or nearby location. These models however do not account for the situation where the activity is secondary and subservient to a selected main destination that is different from the shopper's origin.

Finally, existing spatial models do not account for the different strategies that decision makers may employ (Swait and Adamowicz 2001), which in our case will manifest as differences in attribute sensitivities for different trip conditions. In the context of refueling, Rossi and Chintagunta (2016) identify three segments of consumers with different price sensitivities but they do not distinguish between different trip conditions. Differences in trip conditions have been found to be a greater determinant of consumer differences than traditional segmentation variables such as demographics or type of good (Heider and Moeller 2012). As will be discussed, they also result in temporal variation in spatial competition.

*MODEL SPECIFICATION*

Our model is an extended discrete choice model (Ben-Akiva and Lerman 1985). Suppose a situation where a consumer $i$ will need a convenience good (e.g., gasoline) during a trip from origin $O_i$ to main destination $D_i$. The consumer may become aware of the need while at the origin of the trip or during the trip, at a particular time and location $X_i$. The awareness invokes a decision strategy $s$ from set of strategies $S$. The probability of $i$ selecting outlet (gas station) $j$ under strategy $s$ is

$$P^i(j|s \in S) = \frac{\exp\left(V(X_i)_j^i\right)}{\sum_{j' \epsilon C^i} \exp\left(V(X_i)_{j'}^i\right)} \quad (1),$$

where $V(X_i)_j^i$ is the systematic utility of gas station $j$ observed by consumer $i$ currently located at $X_i$ and $C^i$ is the choice set available to $i$. The probability that strategy $s$ is adopted is

$$Q_{(s \in S|z^i)}^i = \frac{\exp(U_s(\mathbf{z^i}))}{\sum_{s' \epsilon S} \exp(U_{s'}(\mathbf{z^i}))} \quad (2),$$

where $U_s(\mathbf{z^i})$ is the utility of strategy $s$ as a function of individual $i's$ trip characteristic $\mathbf{z^i}$. Combining (1) and (2), the probability that outlet $j$ is selected is (Swait and Adamowicz 2001)

$$P_j^i = \sum_{s \epsilon S} P^i(j|s \in S) \cdot Q^i_{(s \in S|\mathbf{z^i})} \quad (3).$$

The systematic utility of gas station $j$ $V(X_i)_j^i$ accounts for its non-spatial attractiveness as well for the spatial separation from the consumer's position. The representation of the non-spatial attractiveness in our model is not principally different from any other retail model. Typically, retail models measure non-spatial attractiveness as floor size, product assortment, brand image, price level, or any other alternative-specific scalar attribute (Newing et al. 2015; Popkowski-Leszczyc, Sinha and Timmermans 2000; Weisbrod et al. 1984). In refueling the quality of service has also been found to be important, after location and price (Plummer, Haining and Sheppard 1998). Similar to Houde (2012), we use a composite index ($QUAL_j$) to operationally represent the attractiveness of the gas stations, as described in the data section.

To represent spatial attraction, two aspects must be accounted for: spatial separation and spatial structure (Fotheringham 1991). The spatial separation between the individual and an alternative reflects the cost, effort and time required to access it (Bell, Teck-Hua and Tang 1998), including the perceived convenience of access (Bellenger and Korgaonkar 1980). Spatial models, including models regarding refueling (e.g Chan et al. 2007), typically capture separation by some measure of the direct distance between the chooser and the alternative. In on-the-way convenience settings such as refueling, this factor however takes on a special meaning and should account for the deviation from the traveler's path to the final destination (Claycombe 1991; Houde 2012). Descriptive studies of motorists' refueling behavior by Kelley and Kuby (2013) and Kuby et al. (2013) provide evidence that consumers perceive the locational convenience of an outlet as requiring the "least deviation" from the traveler's original path as well as in "being closest" in terms of direct distance from the origin.

Our model hence accounts for spatial separation by including a distance as well as a detour component. The direct distance ($DIRECT_{ij}$) between chooser $i$ and outlet $j$ is the shortest path from the location of $i$ ($X_i$) to alternative $j$ located at $X_j$, i.e., $d_{X_i - X_j}$. Instead, the detour is the incremental distance, relative to the shortest route to the main destination, required to reach that alternative (Eaton and Tweedle 2012; Houde 2012; Pramono and Oppewal 2012; Sun et al. 2016). Hence, it is the extra distance travelled when deviating from the shortest path to destination $D_i$ via $j$. We scale this incremental distance with the remaining distance between the current position and the destination to be able to independently estimate its relative impact:

$$DETOUR_{ij} = (d_{X_i - X_j} + d_{X_j - D_i} - d_{X_i - D_i})/d_{X_i - D_i} \qquad (4).$$

If $i$ is aware of the purchase need before the trip commences the location of the individual is set to the origin $X_i = (x_{origin}, y_{origin})$. Otherwise, the location is set as the point where the individual became aware of the purchase need (point of awareness – $X(A)_i$), in this case $X(A)_i = (x_i, y_i)$, derived using trip information as explained in the data collection section.

Spatial structure concerns the impact of the spatial configuration of the alternatives, in particular local competition and agglomeration effects. Local competition is typically represented by the number of neighboring alternatives within a certain geographical range as determined by, for example, a threshold distance (Eckert and West 2006; Netz and Taylor 2002; Nishida and Remer 2018), street adjacency (Chandra and Tappata 2011), within the same route (Rossi and Chintagunta 2016) or administrative territory (Iyer and Seetharaman 2008; Sinha 2000). Following Eckart and West (2006), we define the level of local competition ($COMP_j$) for alternative $j$ as the number of local competitors within a half kilometer distance from $j$ ($n_k$).

The current model includes agglomeration through an accessibility index that measures, for each outlet, its potential for spatial interaction with other activities in the surrounding zonal areas. As proposed in the literature, the level of spatial agglomeration affects a consumer's patronage decisions as it provides opportunities for multipurpose shopping (Arentze et al. 2005; Bucklin 1963; Dellaert et al. 1998; Fotheringham 1988; Oppewal and Holyoake 2004) or because, as found for refueling, outlets in busy commercial areas are more visible or identifiable (Dingemans, Sperling and Kitamura 1986). The index describes the spatial distribution of possible activities and the ease of reaching those destinations from the particular outlet. We formulate the agglomeration effect for outlet $j$ ($AGGL_j$) as in Bhat et al. (2002):

$$AGGL_j = \sum_{l \in L} OPP_l \cdot \exp\left[\frac{(r_{jl}/T^*)^2}{-2}\right] \quad (5)$$

where $OPP_l$ is the number of retail opportunities in zone $l$ (element of the set of zones $L$), as derived from local census data, located $r_{jl}$ minutes away from gas station $j$, and $T^*$ is the travel time for the relevant purpose (work or non-work; our study uses the observed sample average).

In sum, our model specifies utility $V(X_i)_j^i$ in equation (1) as:

$$V(X_i)_j^i = \beta_1 DETOUR_{ij} + \beta_2 DIRECT_{ij} + \beta_3 COMP_j + \beta_4 AGGL_j + \beta_5 QUAL_j + e_{jt}^i \qquad (6),$$

where error term $e_{jt}^i$ is independently identically distributed based on an extreme value distribution and $\beta_1$ to $\beta_5$ are coefficients to be estimated. This error distribution implies that we assume there is no spatial autocorrelation, that is, there will be no violation of the Independence of Irrelevant Alternatives (IIA) assumption. Spatial autocorrelation exists when data are related due to their geographical location (Dubin 1988), manifesting the *First law of geography* that "everything is related to everything else but near things are more related than distant things" (Tobler 1970) . As a consequence of spatial autocorrelation, utilities of outlets that are close to each other may be correlated due to some common (but not fully observed) factors (Bhat and Guo 2004). In our model, spatial structure variables $COMP_j$ and $AGGL_j$ are assumed to capture the most significant spatial factors, in particular local competition and surrounding spatial opportunities. However, there may also be unobserved spatial correlation in the non-spatial qualities, for example it is possible that due to differences in competition intensity, outlets in more clustered locations have higher $QUAL_j$ values than spatially monopolist stations. Our study therefore includes the application of Moran's-I test for spatial autocorrelation (Li, Calder and Cressie 2007; Moran 1950) to the Quality scores ($QUAL_j$) in our dataset.

Finally, following Kitamura and Sperling (1987) we propose that the key characteristics of a trip determining a consumer's choice strategy when refueling are the type of activity and the familiarity with (or 'mental map' of) the refueling options for that trip. To represent these characteristics, we include three dummy variables: trip regularity ($REG_i$: 1 the route is travelled at least three times per week, 0 otherwise), need awareness ($AWA_i$: 1 the consumer was aware before commencing the trip that there would a need to refuel; 0 otherwise), and the time of day ($MOR_i$: 1 if the trip commenced between 6 and 11 am; 0 otherwise).

We can now specify the utility of strategy $s$ for individual $i$ based on $i$'s trip characteristics in vector $\mathbf{z^i}$ in the Latent Strategy model (equation 3) as:

$$U_s(\mathbf{z^i}) = \alpha_0 + \alpha_1(REG)_i + \alpha_2(AWA)_i + \alpha_3(MOR)_i + \alpha_4(REG)_i(AWA)_i(MOR)_i + \mu_{is} \quad (7)$$

where $\mu_{is}$ is independently distributed according to a type-I extreme value distribution and coefficients $\alpha_0$ to $\alpha_4$ are to be estimated. Interaction term $(REG)_i(AWA)_i(MOR)_i$ represents that commuters have a higher route choice efficiency in the morning (Ta, Zhao and Chai 2016).

*DATA COLLECTION*

In a major city in Asia (pop. 1.8 million, 200 km$^2$), 287 motorists were intercepted at seven major gas stations across the city. The total urban area consists of 241 zip code areas and has 72 gas stations (Figure 1). The seven locations include residential areas, areas with high levels of retail activity, and commuter roads[1]. Refueling motorists were approached on weekdays to obtain the origin and destination location of their trip. Using suburb names and important landmarks (inter-city travelers were excluded), these were later projected onto zip code centroid points to obtain coordinates $(x_{origin}, y_{origin})$ and $(x_{dest}, y_{dest})$. Motorists also indicated their trip purpose (work or non-work), regularity (whether they travelled the route at least three times a week) and when they had become aware of the need to refuel (before or during their trip). If the latter, they estimated how many minutes before entering the gas station they had become aware, which allowed deriving the point of awareness $(x_t, y_t)$. Over-the-road distances between zip codes were obtained through a GIS analysis. Further details are in Appendix A.

[1] Choice-based sampling is commonly used in choice modeling especially in spatial contexts (e.g., Basar and Bhat 2004; Haab and Hicks 1997)

INSERT FIGURE  1 HERE

The majority of intercepted drivers was male (82.4%) and 64.4% indicated that the main purpose of their trip was work or study related (including commuting). 23.7 % came from within 1 km of the gas station while the majority (55.7 %) came from origins located between 2- 5 km from the visited station. This high proportion of short distances is due to the congestion in the study area--the distances equate to travel times between 9 and 21 minutes. The remaining 20.6% came from further than 5 km. These findings suggest that the gas stations depend mostly on passing traffic (Goodchild and Noronha 1987; Houde 2012; Nicholas 2010). The average shortest distance from origin to main destination was 5.9 km (s.d. 4.0). While 70.7% had been aware of the pending need to refuel before leaving their origin, those who became aware during their trip reported 9.6 mins (s.d. 7.3) on average as the time between becoming aware and entering the station. Seven respondents (2.4%) were excluded as they indicated refueling was the main purpose of their trip. This leaves 280 respondents.

The tank levels when motorists entered the gas station were also recorded. 29.0% had the fuel gauge at less than 1/8th, 28.2 % had the gauge between 1/8th and 1/4th and 42.8% had more than 1/4th fuel capacity left before refueling. This indicates refueling motives involved more than just 'running out of fuel' (cf. Kitamura and Sperling 1987). Quality index scores were obtained for all 72 fuel stations from an independent appraisal organization who evaluates the quality of service, quantity assurance, facility maintenance, retail format and product offering. We use it as a proxy for the non-spatial attractiveness (QUAL) in equation (6); its mean was 86.24 (s.d. 4.19).

*MODEL ESTIMATION AND COMPARISONS*

We used Maximum Likelihood estimation. The log-likelihood function for equation (3) is:

$$LL((\boldsymbol{\alpha}, \boldsymbol{\beta})|\boldsymbol{Z}, \boldsymbol{X}) = \sum_{i=1}^{I} \sum_{j=1}^{N} Y_{ij} \ln \sum_{s=1}^{m} P(\boldsymbol{\alpha}, \boldsymbol{X})_{(j|s \in S)}^{i} \cdot Q(\boldsymbol{\beta}, \boldsymbol{Z})_{(s|z^i)}^{i} \qquad (8)$$

where $i$ is the number of cases and $Y_{ij}$ is set to 1 if consumer $i$ refueled at alternative $j$ and 0 otherwise. $\boldsymbol{X}$ represents the station attributes and $\boldsymbol{Z}$ the trip characteristics as described earlier. We estimated coefficients $\boldsymbol{\alpha}$ and $\boldsymbol{\beta}$ for one to three strategies; the two-strategy model had the best fit (AIC: 1499.70, 1392.40, 1416.60; BIC: 1533.20, 1519.30, 1590.7 for $n_S$=1, 2, 3). As explained earlier, to test for possible spatial autocorrelation, we apply Moran's-I test (Li et al. 2007; Moran 1950) to the Quality scores ($QUAL_j$). This test whether the stations' Quality scores can be assumed to be randomly distributed failed to reject the randomness null hypothesis (I(Observed) = -0.037 vs I(Expected) = -0.014; p=.79, n.s.) and we thus conclude there is no autocorrelation and the application of the logit specification in equation (1) is valid.

We next compare the two-strategy model with four benchmark models (Table 1). The first is the single strategy model. Its coefficients reveal the merits of modeling the strategy variations. The second benchmark model is a gravity-based model with local clustering ($COMP_j$) and agglomeration ($AGGL_j$) effects as formulated in (5) and (6), in parallel with the models by Borgers and Timmermans (1987) and Fotheringham (1988). We also include a 'pure' gravity model. This only models the trade-off between an outlet's attractiveness and the direct distance between the store and the chooser (Huff 1964), without reference to the origin-destination path, and assumes the chooser is at the origin. These are the assumptions typically made in spatial modeling and refueling studies (e.g., Chan et al. 2007). We finally also estimated a mixed model with an equivalent specification as the single strategy model but including random coefficients for the five utility predictors, thus capturing heterogeneity without specifying a latent structure.

INSERT TABLE 1 HERE

Table 1 shows that the latent strategy model significantly better fits the sample data than the comparison models. Looking at the gravity-based benchmark models, distance and site quality are significant and have the expected signs. None of the spatial structure terms are significant in the extended gravity model, suggesting these effects do not apply in the present context. The single strategy model shows a large negative effect of detour in addition to a significant effect of distance. The effect of quality is of comparable size in the different models. The mixed model reveals significant heterogeneity for distance and detour, and a fixed effect for quality, but no other significant effects. In contrast, the latent strategy model has significant effects for all five utility predictors and for three predictors in the strategy selection component.

*FINDINGS FROM THE LATENT STRATEGY MODEL*

The latent strategy model shows different behavioral effects for each strategy. The first strategy is consistent with the typical gravity model, with direct distance displaying a decay effect. Because of the consumer's greater focus on the immediate need and immediate environment, we call this strategy "immediacy-oriented". The second strategy reveals effects that are inconsistent with traditional retail spatial theory. It presents a type of spatial behavior in which the choice is more defined by the travel path to the destination, as demonstrated by the large effect for the detour parameter, the *positive* direct distance decay and the significant spatial structure effect. We refer to this strategy as "destination-oriented" because the consumer puts a greater focus on the destination and is less willing to deviate from the main travel path.

*Strategy-specific Effects versus Traditional Retail Models*

The positive distance decay effect in the destination-oriented strategy is remarkable as it is inconsistent with typical retail gravity-based theory, which is based on the main assumption that the farther places, people and activities are apart, the less they interact (Haynes and Fotheringham 1984; Tobler 1970). A possible explanation for the effect is that motorists engaging in this strategy perceive a higher trip efficiency when the refueling can be conducted near the destination (Ta et al. 2016), leaving the main part of their trip without disruptions from extra maneuverings. Another explanation is that these motorists display a locational bias towards their own activity space due to their greater knowledge of the area (Johnston 1972). The coefficients in the strategy selection component are consistent with this interpretation. They reveal that morning commuters and those who are early aware of the need to refuel engage more in the destination-oriented strategy. These consumers may have better knowledge of the refueling locations near their destination (Dingemans et al. 1986).

A second notable finding is the significance of the local clustering and agglomeration effects in the destination-oriented strategy; these were not significant in the immediacy-oriented strategy or in the benchmark models. The negative coefficient for local clustering indicates that the nearby presence of another gas station reduces a gas station's choice probability. This confirms the finding from Netz and Taylor (2002) regarding the push for gas stations to spatially differentiate to maintain their market share. The positive coefficient for agglomeration indicates that accessibility to other activities positively influences a station's choice probability. This finding confirms that agglomeration affords consumers benefits through the opportunity to conduct multipurpose activities (Brown 1989; Dellaert et al. 1998; Fox and Sethuraman 2010). The insight from our model is that these effects only exist when a destination-based strategy is adopted.

*Decision Strategy Adoption*

To describe the decision strategy probabilities, we apply the latent strategy model in equation (3) to each possible combination of the values of the trip variables (i.e., route regularity, time of awareness and time of day). The results (Figure 2) show that motorists are overall most likely to adopt an immediacy-oriented strategy. However, when they are aware of the pending need before departing, there is a significant tendency to adopt a destination-oriented strategy, which implies a greater likelihood that a refueling location nearer to the destination is selected.

INSERT FIGURE 2 HERE

Motorists who become only aware during their trip are more destination-oriented when traveling along a non-regular route. Possibly their lower familiarity induces unassisted spatial sequential search processes (Dellaert and Häubl 2012), resulting in the inspection of alternatives further on the route. The model indicates these motorists are more likely to select alternatives closer to their destination as they are unlikely to return to a previous alternative.

*ILLUSTRATIVE MODEL APPLICATIONS*

We present three illustrations of how our model can be useful. We first apply the model to explore possibilities for segmentation in our study area. Next, we apply the model to assess how competition levels in our study area vary by location and time. Finally, we present different scenario of abstracted duopolistic spatial competition settings to demonstrate the effects of our model variables in a more general sense.

*Segmentation Opportunities*

The identified decision strategies may provide a basis for segmentation and targeting for convenience-based retail and services operations. In particular, our analysis suggests that those who have adopted a destination-orientated strategy are more open to considering combining their stop with other activities in the same area and also are more likely to be aware of their need to purchase (refuel) before they commence their trip. This means these consumers could be approached earlier during their trip, that is, while still at their origin location, to remind them of the forthcoming need, including possibly reminding them of a possible need to purchase additional, non-fuel, items and plan their trip accordingly. Similarly, if another store type, for example a supermarket, were co-located with a fuel station a reminder of its fuel offer might result in a visit to this store. Note that in terms of our model, this may mean this store becomes a main destination (prior to the final destination) and so, when using the model for predictive purposes the trip should be respecified accordingly.

In order to implement a segmentation based on decision strategies a retailer will have to assess whether a consumer is destination-oriented or immediacy-oriented. In the present study context, the two main identifying variables are whether the consumer is pre-aware and whether the trip is a commute. Both relate to a higher probability of adopting a destination-oriented strategy. Therefore, if a retailer can identify which consumers are pre-aware or which trips are more likely to be a commute, the retailer, or the affiliated store type, can seek to communicate its presence or its specific product offer to these consumers in order to stimulate them to schedule their trip to include this retailer's outlet. In particular, if a retailer is located further away from the consumer's origin and there are competing retailers located between the origin and the retailer on the consumer's path, such a strategy may increase the probability of a visit.

It may not always be easy or possible to infer where particular target consumers are located or at what time they are most likely to travel. However, there may be diagnostic variables that can help identify and predict a customer's most likely adopted strategy. To illustrate how the model can predict differences between existing classifications of consumers, that is, for variables not included in the model specification but that are easily observable, we derive the strategy selection probabilities for each respondent in our sample and then analyze how these differ for possible segment classifiers.

First, we explore gender as a possible segmentation base. While the results reveal a slightly lower chance of selecting a destination-oriented strategy for males than for females (29.0% vs 31.5 %), this difference is not statistically significant, which is possibly due to the overrepresentation of males in our sample. Accordingly, including it as an additional latent strategy predictor neither showed any significant effect. However, as an illustration, if it had been significant, this effect would indicate that the females in our sample were destination-oriented and so may be targeted in different ways than males.

Next, looking at tank level our analysis suggests this is a potential basis for segmentation[2] as different tank levels as observed at the refueling station differ significantly in their probability of having adopted a certain strategy ($F_{(2,277)} = 3.91$, p<.05). Those who arrived with an almost empty tank (<1/8th full) have a higher probability of having selected a destination-oriented strategy (32.9%) than those with between 1/8 and ¼th left (28.7%) or those having more than ¼th capacity left (26.3%). This difference suggests that destination-oriented consumers traveled longer before refilling their tank, possibly because they had more carefully planned where to refuel. These consumers are thus likely to have the greater transaction amounts and from a revenue-generating perspective would be more relevant to target. Given that the tank level is observable at the point of purchase it provides an opportunity for an on-the-spot diagnosis of a customer's likely decision strategy, which next could inform promotional and cross selling activities.

---

[2] Because the tank level was observed at the point of refueling, not at the point of awareness, the tank level logically could not be included as a determinant of strategy selection in the model.

In order to explore differences in the customer base of existing stations in our area we next apply the model to derive the distribution of strategies for each of our seven sampled stations. The adoption probabilities differ significantly between the stations ($F$ (6,273) = 6.41, p<.001). The station with the highest percentage of destination-oriented customers (marked A in Figure 1), at 38.5%, is located near a freeway entrance and close to a major industrial area. This means it will have many commuters among its customers and consumers who will plan ahead because of entering the freeway, which aligns with our model findings. In contrast, the station with the lowest percentage of destination-oriented customers (23.0%; marked B in Figure 1) is located near the CBD and is surrounded by competing stations, and so is a location where consumers can expect to find a station whenever needed, which is consistent with the relatively high percentage of immediacy-oriented customers. These two stations could therefore adopt different strategies for reaching their key customers.

*Measuring Local Competition Variation Within a Local Market*

We next demonstrate how the model more generally can provide insight into the level of spatio-temporal competition between stations in a local market as represented by a sample of consumers. At the individual level, the substitutability between two spatially distributed outlets $j$ and $j'$ is modeled as the change in the choice probability of $j$ for a unit change in the utility of $j'$ as observed by individual consumer $i$ ($\gamma_{j'-j}^{i}$). As noted by Sethuraman, Srinivasan and Kim (1999), under the logit model the substitutabilities are symmetric ($\gamma_{j'-j}^{i} = \gamma_{j-j'}^{i}$). At the local market level, with $I$ spatially distributed customers[3], the substitutability can be calculated as (Ellickson et al. 2020)

$$\gamma_{j'-j} = \frac{1}{I}\sum_{i=1}^{I}\frac{\partial P_j}{\partial U_{j'}} \quad (9).$$

Using partial derivatives of equations (1) - (3), this can be rewritten as

$$\gamma_{j'-j} = -\frac{1}{I}\left\{\sum_{i=1}^{I} Q_1^i \left[P^i{}_{(j|s=1)} \cdot P^i{}_{(j'|s=1)}\right] + (1 - Q_1^i)\left[P^i{}_{(j|s=2)} \cdot P^i{}_{(j'|s=2)}\right]\right\}(10).$$

As the latent strategy component $Q$ is a function of whether a trip is made in the morning or afternoon, the model captures the temporal variation in competition between $j$ and $j'$.

To illustrate, we measure the competition among a subset of nine stations in our study area, all located in a two-by-two km area NW of the CBD as shown in Figure 1. The results in Table 2 reveal that the competition in the afternoon (upper diagonal) is much more intense, on average four times as high, than in the morning. They also show that the strongest competitor is not always the closest one. For example, whereas stations 3 and 6 are just 400 meters apart on the same road, they compete very little in the morning. In contrast, in the afternoon they show a very high level of competition. The lower levels of competition in the morning in our area can be attributed to a greater likelihood of adopting a destination-oriented strategy, which manifests in more restricted travel paths and a greater decisiveness regarding where to refuel. In sum, the analysis reveals the spatio-temporal variation in competition among gas stations and how this depends on consumer decision strategies in addition to the stations' non-spatial attractiveness and spatial distribution.

INSERT TABLE 2 HERE

---

[3] While the outlets are locally clustered, the consumers can originate from any area within the city.

*Duopolistic Scenarios Demonstrating the Effects of Various Spatial Factors*

We now demonstrate how our model factors play out in general and how they impact the spatial distribution of demand in various competitive scenarios. Our model effects revealed that the impact of the direct distance to an outlet can be minimal or can be even positive, depending on the individual's decision strategy. This demonstrates that the distance-decay effect assumed in conventional spatial choice models can be diluted and overridden by other, stronger spatial factors such as the individual's movement path, knowledge (mental map) of the area, or the interest to conduct multipurpose activities. It can even mean that, all else equal, an outlet located farther away in terms of direct distance is selected over one that is nearer to the consumer's current location.

To demonstrate these effects we present five duopolistic simulation scenarios. They illustrate the effects for point of awareness (scenarios 1 and 2), agglomeration (scenarios 3 and 4) and quality differences (scenario 5). For each scenario, we map how a consumer's probability of choosing one of two hypothetical refueling stations varies as the target station's location varies relative to the other station (the competitor), when the latter remains in a fixed location. The simulations are implemented for a hypothetical 100 x 100 grid city with a single route connecting one origin and one destination. The origin (O) is located at grid (x,y) coordinates (20, 50) while the destination (D) is located at coordinates (80, 50), as shown in Figure 2. A single consumer is simulated to be traveling from O to D. The competitor station remains fixed at the middle of the path (50, 50), while the location of the target station is varied across all nodes of the grid.  For all scenarios, the time of day is fixed as *morning* and the route is always assumed to be *regular*. The non-spatial attractiveness (i.e., the quality score) is equal for both stations except in scenario 5. The choice probability for the target station is derived from the model equations for every node of the grid as a possible competitive location. The density of the probability values is mapped to show the spatial distribution of the simulated consumer's probability of choosing the target station.

*Scenario 1: Aware before departure.* Scenario 1 concerns the case where the consumer is aware of the need to refuel before the trip commences, so, at the origin point (O). Figure 3a shows that, as expected, the station is more likely to be chosen if it is located nearer to the origin, given the motorist is still located in the origin when first considering the refueling need. The maximum probability is 55%, at location (24, 50). However further observation also reveals that there is less decline in choice probability between the competitor station and the destination than between the competitor and the origin. Even if the target station is located on the destination point, it will still have a relatively high choice probability (41%). This is in accordance with Kitamura and Sperling (1987) that motorists in a planned refueling mode have a bias towards stations located near their origin or destination.

INSERT FIGURE 3 HERE

*Scenario 2: Aware during trip.* Our second simulation depicts the situation where the need to refuel is realized only after leaving the origin. The point of awareness is assumed to be distributed evenly along the route instead of being fixed at the origin point. The contour map (Figure 3b) reveals the choice probability is initially rather low but then rises towards its maximum of 58% at location (57, 50), shortly after the competitor's location. This pattern reflects how the consumer's total travel costs as resulting from the detour vary with the consumer's location and the variation in choice strategies.

The asymmetrical pattern of probabilities in Figures 3a and 3b indicates that the additional travel needed to reach the gas station away from the intended travel path is strategy specific and not uniform across the travel path. For a motorist who adopted a destination-oriented strategy the travel cost is lower (i.e., the likelihood to deviate is higher) when she is closer to the origin and increasingly higher afterward. The opposite effect is observed for the immediacy-oriented strategy, the likelihood to deviate being higher closer to the destination. This finding signifies the possible bias in classic Hotelling-based duopolistic competition studies, where the cost is assumed fixed and uniform across the travel path (Eaton and Tweedle 2012; Martín-Herrán and Sigué 2019). The existence of such asymmetries has however been hinted in descriptive studies of refueling behavior which considered the effects of mental maps and decision modes (Dingemans et al. 1986; Kitamura and Sperling 1987) and the characteristics of urban traffic and trip destinations (Cooper and Jones 2007; Themido, Quintino and Leitao 1998). Inspection of Figure 3b further reveals that the target gas station can gain on the competitor by locating close to it but slightly nearer to the destination. In our dynamic market setting this means there is a significant incentive for the competitor to also move nearer to the destination, which in turn will induce the target to move even further in the direction of the destination.

To better understand the spatial competition between the two stations in this scenario and identify where the equilibrium state will be achieved, we conducted a sensitivity analysis. While varying the locational configuration of *both* stations along the straight path between the origin and destination (i.e., by varying the x coordinates while fixing their y coordinates at 50), we recalculated for each locational configuration the choice probability of the competing station using equation 3. The resulting equilibrium location is at (65, 50). At this point both stations share the same location along the path from origin to destination.

Intriguingly, in the first half of the path, the fuel station located closer to the origin, which hence will be encountered earlier in the trip, will end-up having a lower choice probability. This is against the intuition of spatial dominance, which is that an earlier encountered station will be preferred (Cascetta and Papola 2009). The effect can be attributed to two factors. First, as indicated by the significant negative detour effect in the model, consumers are unlikely to make a u-turn to choose a previously passed outlet. Second, as was shown in Figure 2, even when becoming aware only during the trip, in our setting there is an up to 20% probability that consumers engage in a destination-oriented strategy, which displays an extremely large negative detour factor and a positive coefficient for direct distance. These factors together result in an alternative closer to their destination being preferred over an outlet closer to the origin or current position.

*Scenarios 3 and 4: Agglomeration effects.* Scenarios 3 and 4 replicate the earlier scenarios but add a shopping center at a location away from the travel path to the main destination, at location (50, 75). This location represents a substantial detour but provides the possibility of combining the refueling with additional activities at the center location. In our model, the agglomeration effect was substantial and significant for the destination-oriented strategy but only small and not significant for the immediacy-oriented strategy. Applying the model to the situation where the consumer is aware of the need before departing shows how locations near the shopping center become more likely to be selected. Locations near the origin on the path to this center also benefit substantially (Figure 3c). In contrast, when the consumer becomes only aware during the trip there is no such beneficial effect, the consumer not being able or willing to combine the refueling with the shopping (or other activity) at the center (the map is very similar to 3b, and therefore not shown). This reveals how agglomeration effects depend on the consumers' abilities to plan their trips ahead in order to optimize their travel.

*Scenario 5: Outlet quality.* Scenario 5 shows the effect of an increase in outlet quality—noting quality is only one example of a non-spatial attribute. In a market with price competition the attribute could be substituted with price and similar patterns would be expected to show, although the actual sensitivity of course will be different. To save space we only show the simulation for the case where the consumer is aware before departing. Figure 3d shows the effect of the target station having a 10% higher quality than the competitor (index scores 88 versus 80) and how this affects the choice probabilities. The result is an increase of 10 percent points in probability, with the maximum achievable probability now being 82%, at location (45, 55).

## CONCLUSIONS AND DISCUSSION

### Summary of Findings

This paper addressed the common but neglected situation where consumers need to visit a retail or service outlet while on-the-way to a main destination, which we called on-the-way decision contexts. Refueling is a typical on-the-way context but several other types of convenience shopping equally qualify as on-the-way contexts. On-the-way differs from on-the-go (Heider and Moeller 2012) as it extends beyond food and beverages and involves a particular trip destination, implying a directional component to the valuation of the outlet options.

Our analysis indicated that consumers in on-the-way choice contexts may adopt two different decision strategies. They either employ an *immediacy-oriented* strategy, in which case their priority is to find the most spatially convenient alternative and the decision process is in accordance with the well-known retail gravity principle, where consumers trade-off between store attractiveness and the direct distance between the chooser and the store. Or they adopt a *destination-oriented* strategy, in which case consumers' main concern is their overall trip efficiency. The latter means they place a greater preference on outlets that result in the least deviation from their intended path and that offer greater opportunity for conducting multipurpose activities. Therefore, for them, the attractiveness of an alternative is more affected by its accessibility to other activity locations. It also means that the local competition from nearby alternatives, suggested in the literature (Chandra and Tappata 2011; Cooper and Jones 2007; Kvasnička, Staněk and Krčál 2018; Nishida and Remer 2018), only affects consumers who engage in this strategy, not those who adopted the immediacy-oriented strategy.

We found that the likelihood of adopting a particular strategy relates to specific trip characteristics, in particular trip regularity, the moment the purchase need arises, and the time of day. More specifically, we found that while in our study area the immediacy-oriented strategy is more likely to be adopted, the destination-oriented strategy gains significant probability when the trip is a morning commute and the consumer is aware of the need to purchase before departing. Our results further provide support for the role of familiarity with the local area in directing consumers to particular outlets, as suggested earlier in the literature (Chorus and Timmermans 2010; Dingemans et al. 1986; Johnston 1972).

Our analysis further showed how heterogeneity in consumer choice strategies shapes the substitutability between gas stations in a local market, including strong evidence of spatio-temporal variation in local competition. Our simulations showed how variation in consumer choice strategies impacts the choice pattern and spatial distribution of demand between outlets and cannot be explained by current approaches to retail outlet choice modeling. The latter includes that demand will not always cluster around residential locations (Chan et al. 2007; Iyer and Seetharaman 2008), and is neither evenly distributed within 'narrow' markets (Houde 2012). Instead, the results resonate well with the descriptive findings from Kitamura and Sperling (1987) that spatial demand will be concentrated in and divided between areas around the origin and the destination.

Our analysis also revealed the impact of agglomeration, that is, the extent to which an outlet is clustered with other retail opportunities, on the spatial distribution of demand for convenience-based goods. We showed that for consumers adopting a destination-oriented strategy, outlet choice not only depends on the attributes of the outlet itself but also on its accessibility to other retail opportunities; no such effect existed for those adopting an immediacy-oriented strategy. This finding provides further evidence of the role of agglomeration in retailing as previously found (Arentze et al. 2005; Oppewal and Holyoake 2004; Teller et al. 2016) but also shows how its effects, which are critical for many types of convenience-based retail, are strategy-dependent. Relatedly, in contrast to the retail aggregation principle (Fotheringham 1988; Jensen, Boisson and Larralde 2005), which emphasizes the tendency of similar retail outlets to form clusters, we find that convenience-based outlets such as fuel stations do not benefit from being in a cluster with similar outlets. In fact, similar to Kvasnička et al. (2018) and Netz and Taylor (2002), we find that there is a substantial negative impact of local competition from outlets in the same category. This means there is a strong incentive for spatial differentiation. However, our model shows that this impact is strategy-specific and only occurs when consumers are engaging in a destination-oriented strategy. These findings highlight the difficulties in assessing the impact of local competition at the aggregate level in convenience goods markets (Eaton and Tweedle 2012).

*Contributions*

In terms of contributions to the literature, we firstly clarified an aspect of the broad concept of convenience that hitherto had not been sufficiently defined and provided substantive new insights into the factors that determine consumer outlet choice in convenience-based settings. We identified on-the-way shopping as a particular convenience-based context and provided new insights in the role of location, spatial competition and consumer heterogeneity in this setting, which exists in a range of retail contexts but is particularly well illustrated and relevant for the context of fuel retailing. We proposed and found support in our analysis that consumers in on-the-way choice contexts may adopt two different behavioral decision strategies, each having different effects on the likelihood with which retail outlets are visited and consequently, how outlets compete or complement each other. Our findings next showed that some fundamental assumptions typically used in spatial modeling, such as the assumption of distance decay, are strategy specific and should not be assumed to hold for every spatial choice decision. We also showed that the relevance and impact of agglomeration benefits and local clustering are strategy specific. We further showed how the adoption of a particular strategy depends on the regularity of the trip and the time when the consumer becomes first aware of the purchase need. We finally also contributed to the better understanding of the determinants of spatio-temporal variety in local competition.

Second, we contributed to the marketer's ability to analyze store choices by providing a modeling framework for on-the-way retail outlet choice. The model can be extended to new applications, either in fuel or in other on-the-way choice contexts, to better analyze, map and predict consumer demand and consumer response for a range of spatial settings. The model not only quantifies the effects of various spatial and non-spatial factors in on-the-way choice contexts but also allows simulations of different competitive scenarios. We showed how the proposed model can reproduce and, better than existing models, explain the choice pattern observed in a real on-the-way market and help understand the nature of the spatial asymmetric competition that has been suggested in the literature for a market where the consumers are moving across space when choosing and consuming the product. At the same time, the application of the current model highlighted the limitations of the typical assumptions and approaches taken in published models in dealing with such a spatially distributed market.

*Managerial Implications*

Our proposed approach can be used to help managers better understand their customers, as shown in our analysis for seven sampled stations, and assess the competitive position of their outlets, as shown in the analysis of substitutability patterns across stations in our area. As such it may help find the best location for new outlets that service particular mobile or commuting consumer segments. In addition, it may help policy makers to better assess consumer access to such retail services. Our results also suggest that retail managers operating in on-the-way contexts should consider alternative bases for customer segmentation and targeting. As shown in our analysis of spatial competition, and in accordance with Heider and Moeller (2012), instead of only focusing on demographics, managers could also focus on the time of day and the extent to which consumers are aware of their pending need to purchase. Managers should also be more aware that the benefits of co-location with other retail outlets or other activities will be most relevant if they target consumers who have adopted a destination-based strategy. The latter type of consumers may be of particular interest as they are likely to spend more, since they have further exhausted their resources (e.g., fuel or food).

We also illustrated the possible use of particular diagnostics to infer a customer's likely decision strategy. Fuel retailers may use the pumped amount as a directly observable diagnostic for determining a customer's likely decision strategy which consequently may inform promotional and cross-selling activities at the point of purchase. Our findings suggest that those with a lower tank level are more likely to have adopted a destination-oriented strategy and hence are likely to have traveled longer before refilling their tank as they have more carefully planned where to refuel. These consumers in turn are therefore likely to have the greater transaction amounts, which from a revenue-generating perspective would make them more relevant to target.

In terms of marketing communications, retailers who are in locations that are likely to attract destination-oriented consumers should aim to reach their target consumers early in their journey, so as to make them aware of their pending purchase needs. They could send reminders, or, as is increasingly possible with loyalty cards, use shopper data from other sources such as supermarkets to remind them of the pending need to refuel or restock. Last but not least, our approach could be used to better anticipate demand and competition in novel settings where consumers need to purchase products on-the-way, such as battery recharging for electric cars.

*Limitations and Future Research*

Due to the particulars of our urban context, the present study could not include and assess the effects of brand or price. However, these non-spatial attributes can be included as extra predictors in the utility function when the model would be applied to settings where brand and price do vary (Nishida and Remer 2018; Plummer et al. 1998). Price would be expected to operate in a similar fashion as station quality, although obviously carrying an opposite sign. Quality had a larger effect in the destination-oriented than in the immediacy-oriented strategy in our study. For price it can equally be expected that consumers who adopt a destination-oriented will display a larger effect. This is because these consumers are more likely to be regular travelers on the route and are more likely to plan ahead. They are therefore more likely to be aware of price differences and also more likely to respond to those differences. Such effects in turn may relate also to demographic variables such as income, similar to the findings by Nishida and Remer (2018), such that those with lower incomes are more likely to plan ahead and adopt a destination-oriented strategy.

Regarding the latter, note that Nishida and Remer (2018) could only include income as an aggregate, zonal variable. Instead, our model allows to include income and other sociodemographic variables as individual predictors, hence allows analyzing mixes of customer types across such groups and from across different types of origins and destinations. Similarly, regarding brand effects, these can also be expected to align with the effects we observed for quality, possibly with even stronger effects to the extent that where consumers are brand loyal, or for example hold a loyalty card, they are again more likely to know about and respond to the presence of particular brands, especially when adopting a destination-oriented strategy. We speculate that the reverse may also be the case, that those who are loyal to a particular brand are more likely to adopt a destination-oriented strategy because they are loyal customers to this brand. Moreover, it is quite feasible that when the model is applied to a context where brand and price both vary a larger number of latent strategies can be derived. Hence, for multiple reasons the inclusion of price and brand is a promising area for follow up research. Our analysis demonstrated the usefulness of the approach and how it can be applied to any new sample of observed or intercepted consumers in a local market.

Some further limitations of the current study are that we analyzed the decision when and where to buy, without reference to the decision of what to buy. Further research should account for various aspects of multipurpose shopping within on-the-way shopping contexts. Consumers may stop at a service station to conduct other on-the-way activities, such as purchasing refreshments, withdrawing cash, or as a fill-in grocery shopping trip. Multipurpose shopping can also involve activities such as work, study, or recreation. It is plausible that effects of detour and spatial structure vary with these travel purposes. Other on-the-way contexts than fuel should also be explored, considering also the increasing importance of convenience for supermarkets (Economist 2015).

Our study also included only a limited sample of consumers and as such the empirical findings are limited to this specific survey sample and setting. Our predictor variables had a limited range and some variables did not display any variation and so could not be included, in particular price and brand as noted. It also only involved a limited survey. Follow up research might employ vignette or stated preference and simulation-based methods to overcome these limitations and allow the better separation of location from quality and other station-specific effects as well as better controlling for spatial structure effects (cf. Benoit et al. 2019; Dellaert et al. 1998; Pramono and Oppewal 2012). Future studies should also seek to involve field settings that vary relevant variables and allow the direct use of customer data (cf. Rossi and Chintagunta 2016).

Finally, there may have been survey-based measurement errors, for example in the determination of the precise origin and destination locations. Precise dynamic location information about the motorists was not available for our study but mobile positioning technology no doubt can improve the geometric precision in future applications. Future applications may include studying the effects of geo-positioning systems that advise consumers of their most economical route (Dellaert and Häubl 2012), effects of geo-targeting (Bradlow et al. 2017; Fong, Zheng and Luo 2015), and mobile promotions (Lurie et al. 2018). The current approach can help understand and predict their effects on choice behavior.

APPENDIX A

DATA DESCRIPTIVES

| Variables | Measurement units | Across Sampled Stations | | | | Across All Market Alternatives | | | |
|---|---|---|---|---|---|---|---|---|---|
| | | Mean | SD | Min | Max | Mean | SD | Min | Max |
| **Station Specific** | | | | | | | | | |
| Site Quality | index | 87.61 | 3.27 | 80.54 | 92.93 | 86.24 | 4.21 | 80.00 | 93.36 |
| Average Sales | kiloliters/day | 38.67 | 11.40 | 27.87 | 61.77 | 20.66 | 10.78 | 0.77 | 61.77 |
| **Spatial Variables** | | | | | | | | | |
| Direct Distance to Station | kilometers | 3.23 | 3.07 | 0.03 | 18.02 | 7.74 | 4.22 | 0.03 | 26.62 |
| Detour (relative to Total Length) | - | 0.46 | 1.01 | 0.00 | 8.89 | 3.57 | 6.32 | 0.00 | 90.70 |
| Local Competition | # of competitors | 4.06 | 1.79 | 1 | 7 | 4.22 | 2.25 | 0 | 9 |
| Agglomeration | accessibility index | 1.10 | 0.18 | 0.64 | 1.27 | 1.12 | 0.14 | 0.62 | 1.29 |

| **Individual Specific** | | Mean | SD | Min | Max |
|---|---|---|---|---|---|
| Regularity | 1 if at least monthly (0 otherwise) | 0.84 | | 0 | 1 |
| Trip Length (Origin to Destination) | kilometers | 5.89 | 3.98 | 0.50 | 19.47 |
| Trip Progression | % to final destination | 0.18 | 0.33 | 0.00 | 1.00 |
| Low Tank | 1 if <1/8th (0 otherwise) | 0.72 | | 0 | 1 |
| Time Traveled since aware: | | | | | |
| Aware at origin (n=196) | minutes | 13.22 | 13.04 | 1 | 77 |
| Aware when on their way (n=84) | minutes | 9.63 | 7.31 | 0 | 40 |

TABLE 1

Estimation Results for the Latent (Two) Strategy Model, Single Strategy Model, Extended Gravity Model[1] and Mixed Logit Model.

| Strategy selection component | Latent (Two) Strategy | | | | | | Single Strategy | | | Extended Gravity[1] | | | Mixed Logit | | | | | |
|---|---|---|---|---|---|---|---|---|---|---|---|---|---|---|---|---|---|---|
| | b | sd | t | b | sd | t | b | sd | t | b | sd | t | b | sd | t | b | sd | t |
| Constant | -1.40 | 0.68 | -2.06 | | | | | | | | | | | | | | | |
| Aware (before v during) | 1.30 | 0.54 | 2.40 | | | | na | | | na | | | na | | | na | | |
| Regular (v irregular) | -0.67 | 0.49 | -1.35 | | | | | | | | | | | | | | | |
| Morning (v other times) | -0.47 | 0.41 | -1.16 | | | | | | | | | | | | | | | |
| Morning Commute (=Aware x Regular x Morning) | 1.07 | 0.47 | 2.29 | | | | | | | | | | | | | | | |
| **Outlet selection component** | *Immediacy-oriented strategy* | | | *Destination-oriented strategy* | | | | | | | | | *Fixed effects* | | | *Random effects* | | |
| Detour | -0.34 | 0.12 | -2.83 | -10.84 | 1.61 | -6.74 | -1.15 | 0.13 | -8.47 | na | | | -2.12 | 0.30 | -7.06 | 0.90 | 0.13 | 6.70 |
| Distance | -0.93 | 0.07 | -13.17 | 0.77 | 0.14 | 5.36 | -0.46 | 0.04 | -11.38 | -0.73 | 0.04 | -20.55 | -0.52 | 0.06 | -8.79 | 0.40 | 0.07 | 5.47 |
| Local Clustering | -0.03 | 0.05 | -0.52 | -0.31 | 0.11 | -2.74 | -0.14 | 0.45 | -0.31 | -0.62 | 0.42 | -1.45 | -0.07 | 0.04 | -1.62 | 0.01 | 0.06 | 0.24 |
| Agglomeration | -1.00 | 0.55 | -1.77 | 6.38 | 2.17 | 2.94 | -0.03 | 0.04 | -0.81 | -0.01 | 0.03 | -0.19 | 0.30 | 0.51 | 0.59 | 0.20 | 1.49 | 0.13 |
| Quality | 0.13 | 0.02 | 5.47 | 0.18 | 0.06 | 2.88 | 0.12 | 0.02 | 6.09 | 0.10 | 0.02 | 5.84 | 0.13 | 0.02 | 6.21 | -0.00 | 0.03 | -0.08 |
| Logl at convergence | -680.30 | | | | | | -761.59 | | | -827.70 | | | -739.49 | | | | | |
| Akaike Inf. Crit. (AIC) | 1392.40 | | | | | | 1499.70 | | | 1663.40 | | | 1499.00 | | | | | |
| Bayesian Inf. Crit. (BIC) | 1519.30 | | | | | | 1533.20 | | | 1695.10 | | | 1535.60 | | | | | |

1) For parsimony, the 'Pure' Gravity Model is not shown but it has similar coefficients as the Extended Gravity Model for its only predictors (Distance and Quality).

TABLE 2

Sample-based cross elasticities between nine stations, as derived from the model estimated across all stations and sampled consumers[1,2].

| Station | 1 | 2 | 3 | 4 | 5 | 6 | 7 | 8 | 9 |
|---|---|---|---|---|---|---|---|---|---|
| **1** |  | -0.15 | -0.62 | -0.62 | -0.22 | -0.27 | -0.43 | -0.33 | -0.46 |
| 2 | -0.04 |  | -0.43 | -0.19 | -0.16 | -0.21 | -0.16 | -0.12 | -0.42 |
| **3** | -0.16 | -0.04 |  | -0.45 | -0.57 | -0.83 | -0.45 | -0.37 | -0.78 |
| 4 | -1.14 | -0.04 | -0.17 |  | -0.21 | -0.23 | -0.49 | -0.36 | -0.62 |
| 5 | -0.05 | -0.03 | -0.09 | -0.07 |  | -0.22 | -0.26 | -0.21 | -0.33 |
| 6 | -0.07 | -0.02 | -0.07 | -0.08 | -0.04 |  | -0.21 | -0.16 | -0.41 |
| 7 | -0.38 | -0.04 | -0.12 | -0.48 | -0.10 | -0.06 |  | -0.33 | -0.45 |
| **8** | -0.16 | -0.04 | -0.10 | -0.21 | -0.12 | -0.05 | -0.17 |  | -0.30 |
| 9 | -0.14 | -0.04 | -0.10 | -0.20 | -0.08 | -0.05 | -0.14 | -0.11 |  |

[1]All nine stations are located in a single area NW from the CBD as indicated in Figure 1, three (marked in bold) were in the intercept sample.

[2]Lower diagonal entries are for morning, upper diagonals are afternoon between-station substitutabilities.

FIGURE 1

Survey area with all fuel stations

(A and B are, respectively, the sampled stations with the highest and lowest percentage of destination-oriented customers; inset shows area with nine stations used for substitutability analysis in Table 2).

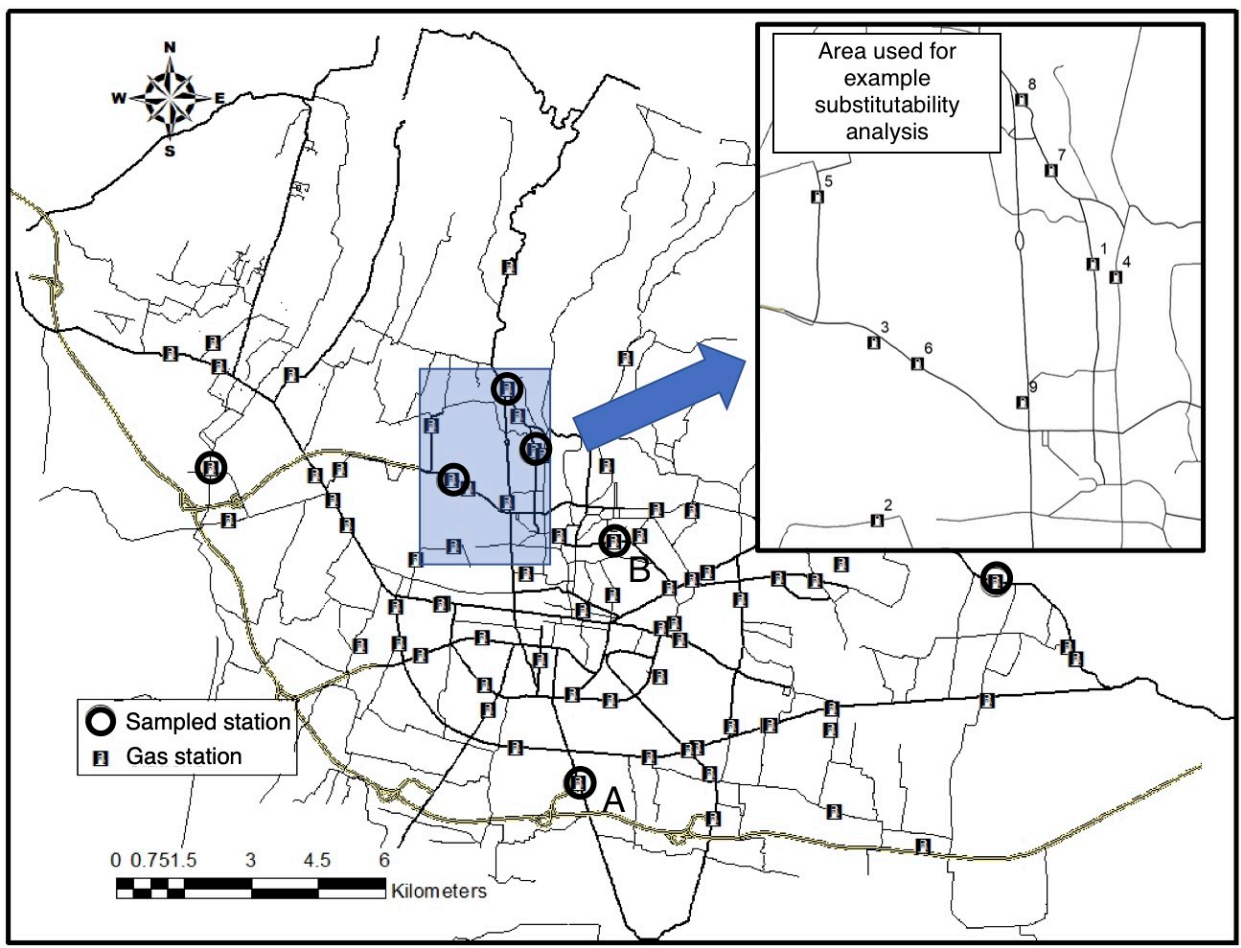

FIGURE 2

Predicted strategy engagement rates for different trip types.

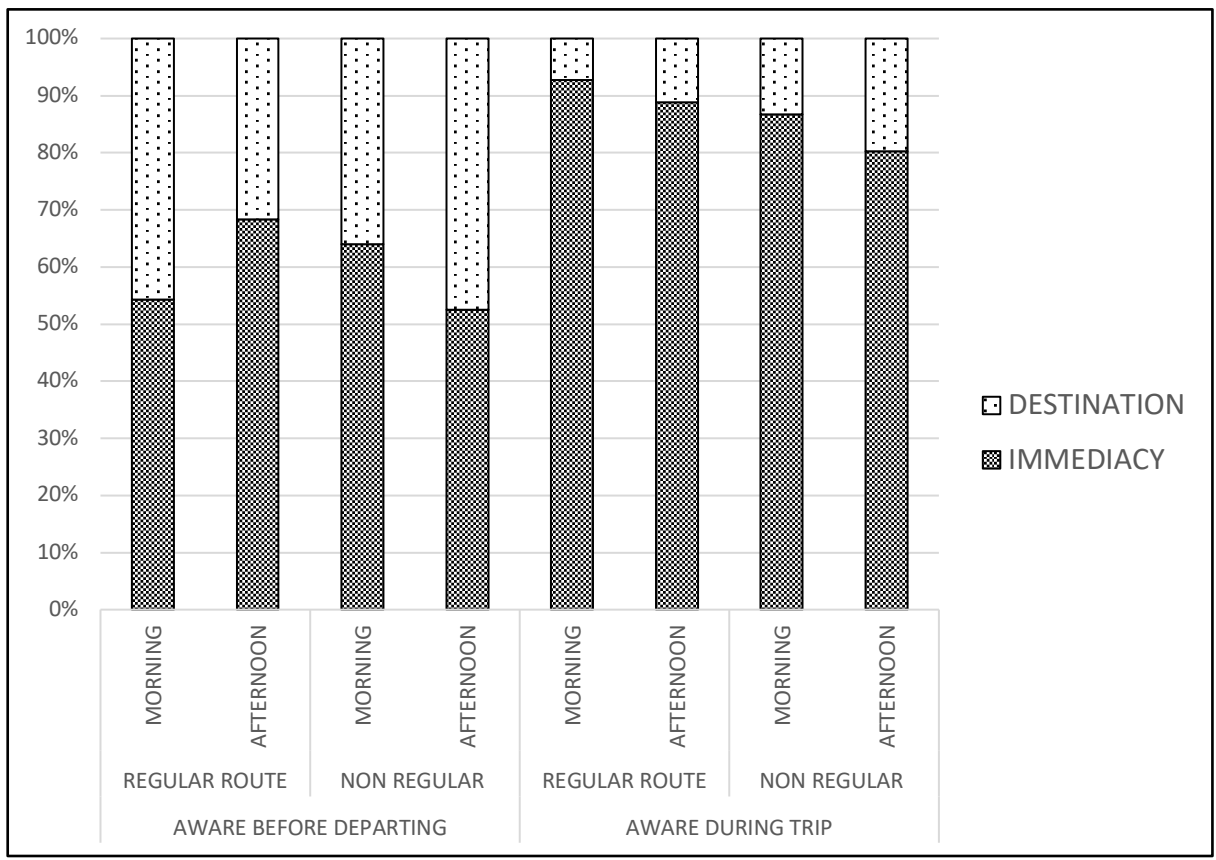

FIGURE 3

Individual choice probability contours for all possible locations of the Target station as the consumer moves from the Origin to the Destination, with one fixed Competitor located in between.

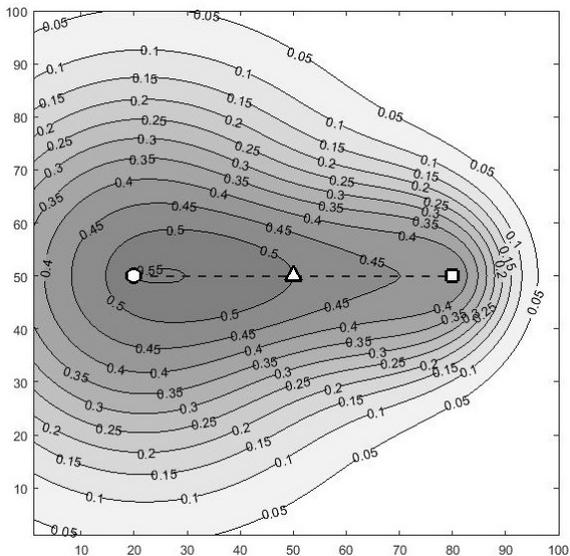

Figure 3a. Aware of need before departing.

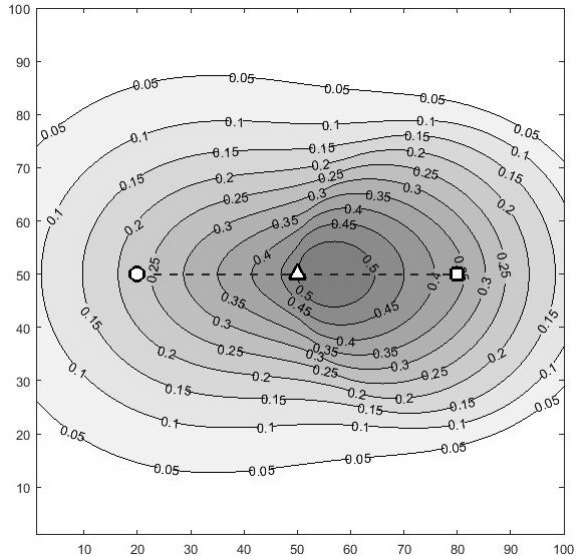

Figure 3b. Aware after departing (point of awareness evenly distributed along the path).

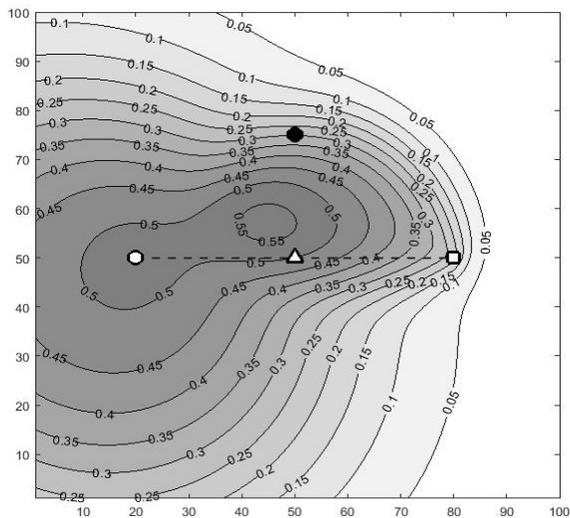

Figure 3c. Aware before departing, with a new shopping center (showing agglomeration effects).

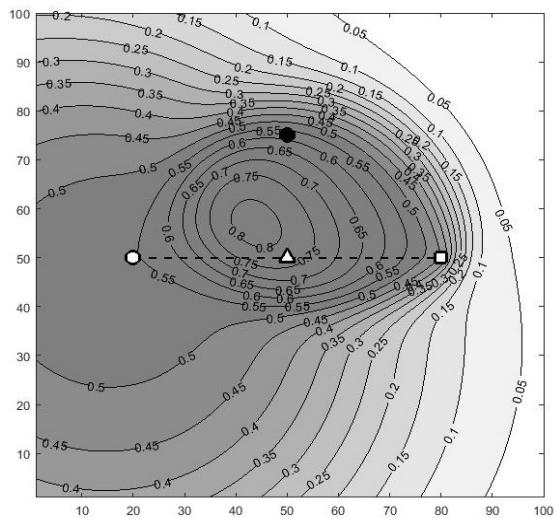

Figure 3d. Aware before departing, with the target being 10% higher quality than the competitor.

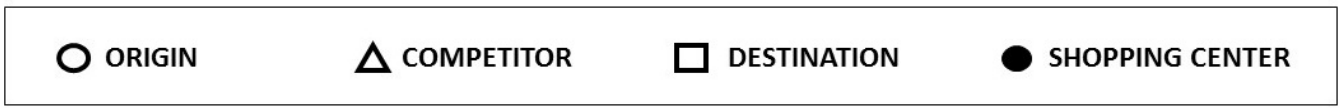